\documentclass[epj,nopacs]{svjour}

\usepackage{comment}
\usepackage{physics}
\usepackage[mathscr]{eucal}
\usepackage{multirow}
\usepackage{graphicx}
\usepackage{amssymb}
\usepackage{amsmath}
\usepackage{subfig}
\usepackage{xcolor}

\newcommand{\be}{\begin{equation}}
\newcommand{\ee}{\end{equation}}
\newcommand{\bea}{\begin{eqnarray}}
\newcommand{\eea}{\end{eqnarray}}

\begin{document}

\title{\centering \boldmath Quintessential inflation in Palatini $F(R,X)$ gravity}

\author{\centering Konstantinos Dimopoulos\inst{1} \and Christian Dioguardi{\inst{2}}$^,${\inst{3}} \and
 Gert H\"utsi\inst{2}\and 
Antonio Racioppi\inst{2}}

\institute{\centering Consortium for Fundamental Physics, Physics Department,
Lancaster University, Lancaster LA1 4YB, United Kingdom \and National Institute of Chemical Physics and Biophysics, R\"avala 10, 10143 Tallinn, Estonia, \and Tallinn University of Technology, Akadeemia tee 23, 12618 Tallinn, Estonia}

\abstract{Palatini $F(R,X)$ gravity, with $X$ the inflaton kinetic term, proved to be a powerful framework for generating asymptotically flat inflaton potentials. Here we show that a quadratic Palatini $F(R,X)$ restores compatibility with the observational data of the Peebles-Vilenkin quintessential inflation model. Moreover, the same can be achieved with an exponential version of the Peebles-Vilenkin potential if embedded in a Palatini $F(R,X)$ of order higher than two. }

\maketitle

\section{Introduction}

The $\Lambda$CDM model represents today the Standard Cosmological Model. The main ingredients of the model are a relativistic matter component (radiation), some non-relativistic matter (cold dark matter, baryonic matter) and a cosmological constant, $\Lambda$, associated with dark energy. The latter drives the current accelerated expansion of the Universe. 

$\Lambda$CDM, while being a simple model, requires extreme fine tuning. The most popular way of improving $\Lambda$CDM is by introducing a scalar field, called quintessence in the literature (e.g. \cite{Copeland:2006wr,Tsujikawa:2013fta} and refs. therein), that in contrast to $\Lambda$ is a dynamical field. If the field varies slowly and has the appropriate energy density today then it can lead to the aforementioned accelerated expansion at present.

At the same time, observations of the cosmic microwave background radiation (CMB) support the idea of a spatially flat and isotropic Universe at large scales, which can be explained by assuming cosmic inflation, i.e. another accelerated expansion of the Universe but during its very early stages \cite{Starobinsky:1980te,Guth:1980zm,Linde:1981mu,Albrecht:1982wi}. 
In its simplest realization the inflation is driven by another scalar field, the inflaton,  with an almost flat potential energy.

Although similar in principle, inflation and dark energy late-time acceleration happen at very different energy scale, respectively around $10^{16}\,$GeV and $10^{-12}\,$GeV. Hence, the two phenomena are usually considered to have different origin. At the same time having an extra \\
quintessence field introduces the coincidence problem: one has to require specific initial conditions so that the currently energy density of the scalar field matches the observed value of the cosmological constant.

Building a model of quintessential inflation overcomes this issue, by realizing the initial condition for quintessence through inflation \cite{Peebles:1998qn}.
However, a working model of the quintessential inflation must satisfy several conditions. 
First of all it has to provide a graceful exit from inflation, providing good predictions for the CMB observables. Second, the scalar field has to survive until present days, this implies that a reheating mechanism that does not rely on its decay has to be provided such as instant preheating \cite{Felder_1999,Campos:2002yk}, curvaton reheating \cite{Feng_2003,BuenoSanchez:2007jxm}, gravitational reheating \cite{grav_reheating,Chun:2009yu,saviour}, Ricci reheating \cite{Dimopoulos:2018wfg,Opferkuch:2019zbd,Bettoni:2021zhq}, reheating by primordial black hole evaporation \cite{Dalianis:2021dbs,RiajulHaque:2023cqe}, warm quintessential inflation \cite{Dimopoulos:2019gpz,Rosa:2019jci}, to name but some. 
Finally, the scalar field potential needs to be very steep in order to bridge the energy gap between inflation and quintessence, allowing for a period of kination, which 
eventually ends due to reheating.
This will eventually lead to the freezing of the scalar field at a value such that its energy density matches the value corresponding to the observed dark energy density.

Several models of quintessential inflation have been built by embedding a scalar field in General Relativity (e.g. \cite{Peebles:1998qn,Dimopoulos:2017zvq,deHaro:2021swo,Bettoni:2021qfs,Jaman:2022bho} and references therein).
On the other hand, Palatini modifications of Einstein gravity proved to be very powerful tools in quintessential inflation model building (e.g. \cite{Dimopoulos:2020pas,Dimopoulos:2022tvn,Dimopoulos:2022rdp,Gialamas:2023flv,TerenteDiaz:2023kgc} and the references therein).
In the standard metric formulation, the only dynamical degree of freedom is the metric tensor, while the affine connection is assumed a priori to be the Levi-Civita one. In the Palatini formulation instead, both the metric and the connection are considered independent dynamical degrees of freedom, and their relation is set by their corresponding equations of motion (EoM). If the action is given by the simple Einstein-Hilbert term, the two formulations are equivalent while in non-minimal theories of gravity they lead to substantially different phenomenological predictions, e.g. \cite{Koivisto:2005yc,Bauer:2008zj}. 

In this article, we are interested in a particular class of non-minimal Palatini models: the $F(R,X)$ models, with $X$ the inflaton kinetic term. Inflation in this class of theories has been extensively studied in \cite{Dioguardi:2023,Dioguardi_2024,dioguardi_2025} as a an extended generalization of the $F(R)$ models already explored in \cite{Dioguardi:2021fmr,Dioguardi_2024}. The $F(R,X)$ models allow to simplify the structure of the scalar field kinetic term in the Einstein frame and consequently heal the dynamical issues of the $F(R)_{>2}$ (that is those containing terms diverging faster than $R^2$) theories out of the slow-roll regime. Remarkably, the $F(R,X)_{>2}$ theories, once the consistency criteria are satisfied, universally provide potential apparently suitable for quintessential inflation \cite{Dioguardi:2023,Dioguardi_2024}.

Quintessential inflation was first introduced in Ref.~\cite{Peebles:1998qn} by Peebles and Vilenkin in order to model both early and late universe expansion 
utilising a single scalar field with scalar potential which behaves as a quartic potential in the early universe, and as an inverse quartic potential at present times. The model successfully reproduces the quintessential inflation behavior however it predicts a tensor-to-scalar ratio $r$ and spectral index $n_s$ which are currently excluded by the current CMB observations \cite{BICEP:2021xfz,Planck2018:inflation}, while also featuring tracker quintessence with too large barotropic parameter. Since this pioneering work, a number of successful quintessential inflation models have been constructed (for recent reviews see Refs.~\cite{Jaman:2022bho,deHaro:2021swo}). More recently, quintessential inflation has been modelled in the context of modified gravity. For example, quintessential inflation models have also been studied in the context of Palatini $F(R)$ gravity for the quadratic choice $F(R)= R +\alpha R^2$. In Ref.~\cite{Dimopoulos:2020pas} the Peebles-Vilenkin potential is discussed,  while in Ref.~\cite{Dimopoulos:2022rdp} the case of the exponential tail is studied in presence of non-minimal coupling with the Ricci scalar. Both models provide the correct behavior for quintessence and predict $r, n_s$ compatible with the current data. 
In this paper we embed the generalized version of the Peebles-Vilenkin potential and the exponential tail in the extended  framework of $F(R,X)$ models and show that in the former case  a viable model for quintessential inflation can be achieved, while the same cannot be achieved with an exponential tail.

In particular, the purpose of this work is to find,  within the Palatini $F(R,X)$ framework,   quintessential inflation setups that describe both inflation in the early Universe and the current accelerated expansion in agreement with the current observational data. While inflation has been extensively studied for this class of models, a description of both early and late universe acceleration is only achieved here for the first time. In particular we restore the compatibility with data of the Peebles-Vilenkin model with inflation and show how the same potential can lead to a dark-energy dominated phase in the late universe. We discuss the whole evolution of the scalar field, compute the predicted inflationary observables and consider a kination period after the inflationary phase, typical of quintessential inflation models. We then consider reheating. Without specifying the reheating mechanism, we constrain the parameter space by taking into account the bounds coming from overproduction of gravitational waves during the kination phase. We finally consider a late-time dark-energy phase, modeled by the tail of the Peebles-Vilenkin potential, and show that it manages to satisfy the coincidence requirement. Morever, we further constrain the parameter space of this class of models by imposing a barotropic parameter within the observational bounds. The full analysis is carried out for both a quadratic choice of $F(R,X)$ and for higher-order models denoted by $F(R,X)_{>2}$.

The paper is organized as follows. In section \ref{sec:Model} we introduce the formalism of Palatini $F(R,X)$ theories. In section \ref{sec:peebles} we consider the quadratic choice and compute the observables of the model for the Peebles-Vilenkin (PV) potential~\cite{Peebles:1998qn}, for inflation, reheating and dark energy, showing the parameter space for which quintessential inflation is viable. At the end of the same section we briefly consider an exponential tail and prove that, while giving very good predictions for the inflation CMB observables, it cannot predict a good quintessential tail for dark energy. In section \ref{sec:peebles_log}, for the sake of completeness, we repeat the same analysis of section~\ref{sec:peebles} for a specific higher-order $F(R,X)_{>2}$ model. We compute once again the observables for inflation, reheating and dark energy by using an exponential version of the Peebles-Vilenkin potential and constrain the parameter space for this model as well, proving that also higher-order $F(R,X)$ can account for quintessential inflation. Finally, in section \ref{sec:conclusions}, we summarize the results of the paper and draw our conclusions.

We use geometric units where $c=\hbar=k_B=1$ and \mbox{$8\pi G=m_P^{-2}=1$}, while the signature of the metric is spacelike.

\section{\boldmath Palatini $F(R,X)$ framework} \label{sec:Model}

The starting point for the Palatini $F(R,X)$ models is the following action:
\be
S = \int d^4x \sqrt{-g^J}\qty(\frac{1}{2} F (R_X) - V(\phi)) \label{eq:S:start} \, ,
\ee
with $F$ is an arbitrary function of its argument, $R_X \equiv R+X$ where $X= - g^{\mu\nu}_J \partial_\mu \phi \,\partial_\nu \phi$ is the scalar field kinetic term and $R = g^{\mu\nu}_J R_{\mu\nu}(\Gamma)$ is the Palatini Ricci scalar built from the metric-independent Ricci tensor, with $\Gamma$ denoting the connection.
We can rewrite the action \eqref{eq:S:start} by introducing an auxiliary field $\zeta$ as follows:
\be
S = \int d^4x\sqrt{-g^J}\qty(\frac{F(\zeta)+F'(\zeta) \qty(R_X -\zeta)}{2} - V(\phi)) \label{eq:S:zeta}. 
\ee
One can easily check that by taking the variation of the action with respect to the auxiliary field $\zeta$, the action in \eqref{eq:S:start} is recovered (given that $F''(\zeta) \neq 0$.)
By means of a conformal transformation $g_{\mu\nu}^E = F'(\zeta) g_{\mu\nu}^J$, we can rewrite the action in the Einstein frame, where the theory is linear in $R$ and minimally coupled to the metric $g_{\mu\nu}^E$:
\be \label{Einstein_action}
S = \int d^4x \sqrt{-g^E} \qty(\frac{R}{2} - \frac{1}{2}g^{\mu\nu}_E \partial_\mu \phi\, \partial_\nu \phi - U(\zeta,\phi)) \, ,
\ee
with
\be \label{eq:einstein_potential}
U(\zeta,\phi) = \frac{V(\phi)}{F'(\zeta)^2} -\frac{F(\zeta)}{2 F'(\zeta)^2} + \frac{\zeta}{2 F'(\zeta)} \, .
\ee
By taking the variation with respect to $\zeta$ we get the corresponding equation of motion:
\be \label{eq:G=V}
G(\zeta) \equiv \frac{1}{4}\qty(2 F(\zeta) - \zeta F'(\zeta)) = V(\phi) \, ,
\ee
which can be solved to get $\zeta(\phi)$. Equation \eqref{eq:G=V} was already introduced in \cite{Dioguardi:2021fmr} where it holds as an approximation valid in the slow-roll regime. However, in this case, eq.~\eqref{eq:G=V} is exact and valid even in the presence of large kinetic terms for the canonical scalar field. In other words, the auxiliary field $\zeta = \zeta(\phi)$ is a function of the canonically normalized field only and not of its derivatives.
By using eq.~\eqref{eq:G=V} one can rewrite $U(\zeta,\phi)$  in eq.~\eqref{eq:einstein_potential} in terms of $\zeta$ only, obtaining
\be
U(\zeta)=\frac{\zeta}{4 F'(\zeta)}. \label{eq:U:zeta}
\ee
This implies that even in the cases for which $G(\zeta)$ cannot be explicitly solved in terms of $\zeta$, it can still be exploited for computing the inflationary observables \cite{Dioguardi:2021fmr}. In all the other (few) cases, $\zeta=\zeta(\phi)$ can be explicitly determined, allowing computations to proceed in the standard way.

In the following, we study two types of $F(\zeta)$ functions. First we consider a quadratic $F(\zeta)$ which has the property to automatically flatten any diverging $V(\phi)$ when $\zeta \to +\infty$ (see eqs. \eqref{eq:G=V} and \eqref{eq:U:zeta}). Then, we study a  $F(\zeta)_{>2}$ i.e. a $F(\zeta)$ of order higher than two. This is a very special case because, first of all, in order to have a viable solution of eq. \eqref{eq:G=V}, $V(\phi)$ must be negative \cite{Dioguardi:2023,Dioguardi_2024}. Secondly, even though $V(\phi) < 0$, the Einstein frame potential $U(\zeta)$ in eq. \eqref{eq:U:zeta} is positive as long as $\zeta, F'(\zeta) >0$. Finally, since $F(\zeta)$ is of order higher than two, $F'(\zeta)$ is of order higher than one, implying that at $\zeta \to +\infty$, $U(\zeta)$ automatically develops a tail approaching zero, which is exactly one of the requirements for quintessence.

\section{\boldmath Quintessential inflation for $F(R_X)=R_X + \alpha R_X^2$}\label{sec:peebles}
This setup has been introduced in the context of fractional attractors in \cite{Dioguardi:2023}. In such a scenario, from eq.~\eqref{eq:G=V} we have $G(\zeta) = \frac14\zeta =V(\phi)$. Hence, by using eq.~\eqref{eq:U:zeta}, we obtain an Einstein frame potential:
\be\label{eq:fractional_potential}
U(\zeta(\phi)) = \frac{\zeta(\phi)}{4(1+2\alpha \zeta(\phi))} = \frac{V(\phi)}{1+8\alpha V(\phi)} \, ,
\ee
where we can immediately see that $U \to 0 \, (\frac{1}{8 \alpha})$ when $V \to 0 \, (+\infty)$. Therefore, in order to have a quintessential tail for the potential we need to consider a form of $V(\phi)$ with a decreasing tail. The simplest example is the Peebles-Vilenkin (PV) potential, the first model made to describe quintessential inflation \cite{Peebles:1998qn}. The Jordan frame potential has then following form:
\be\label{eq:peebles}
V(\phi) =
\begin{cases} 
      \lambda^k (\phi^k + M^k) & \phi\leq 0 \\
      \lambda^k \frac{M^{k+q}}{\phi^q+M^q} & \phi > 0\,, \\
   \end{cases}
\ee
with $k$ an even number. We stress that such a configuration shares a similar form for the Einstein frame scalar potential as in ref.~\cite{Dimopoulos:2020pas}. However, in ref.~\cite{Dimopoulos:2020pas} the scalar field also needs to undergo a canonical normalization, while in our set-up $\phi$ is already canonical normalized. This implies different phenomenological results with respect to ref.~\cite{Dimopoulos:2020pas}.

\subsection{Inflation} \label{sec:inflation}
\begin{figure}[t]
    \centering
    {\includegraphics[width=0.45\textwidth]{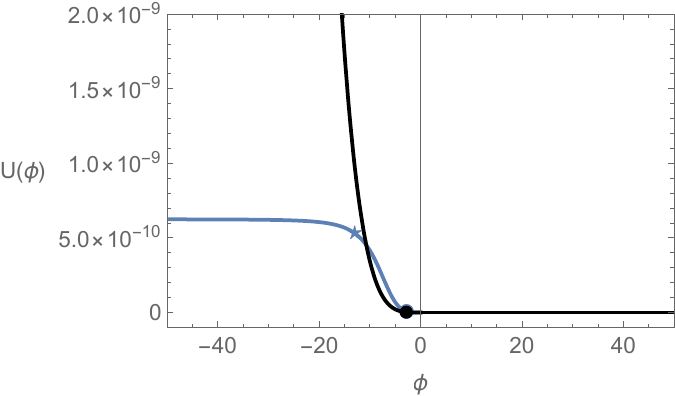}}%

    \caption{The PV potential (black), and the modified PV potential (blue) with $\alpha =2\cdot 10^8$, $q=k=4$, predicting viable CMB observables  $r = 0.017$, $n_s=0.966$ (at $N_e = 60$) and a quintessential tail the solves the coincidence problem at the mass scale $M=1.38\cdot 10^{-13}$. We also show $\phi_N$ (star) and $\phi_{end}$ (dot) in the same color code. $\phi_N$ is not visible for the original PV potential as it lays at $V(\phi_N)\sim 10^{-8}$. We notice that while the potential is modified at the inflation scale, it remains unchanged in the tail. }
    \label{fig:U_peebles_potential}
\end{figure}

The original PV potential does not satisfy the experimental bounds from the CMB observations \cite{BICEP:2021xfz}.
In our case instead, by starting from a Jordan frame potential with the form of the PV potential we obtain an Einstein frame potential with a plateau in the inflationary region, as  shown in Fig. \ref{fig:U_peebles_potential}.
\begin{figure*}[t]
    \centering
    
    \subfloat[]{\includegraphics[width=0.4\textwidth]{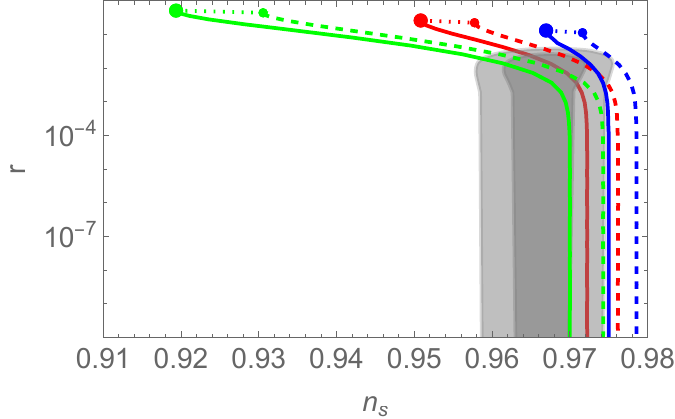}}%
    \qquad
    \subfloat[]{\includegraphics[width=0.385\textwidth]{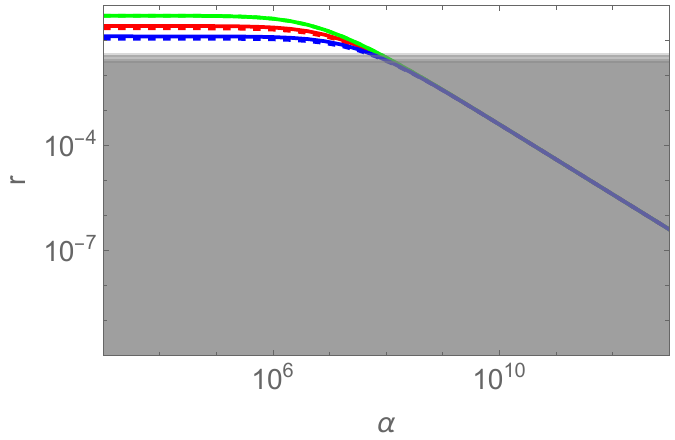}}%
    
    \subfloat[]{\includegraphics[width=0.4\textwidth]{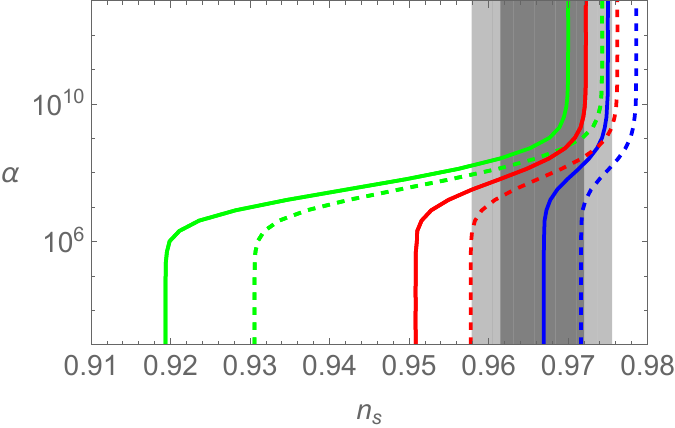}}%
    \qquad
    \subfloat[]{\includegraphics[width=0.39\textwidth]{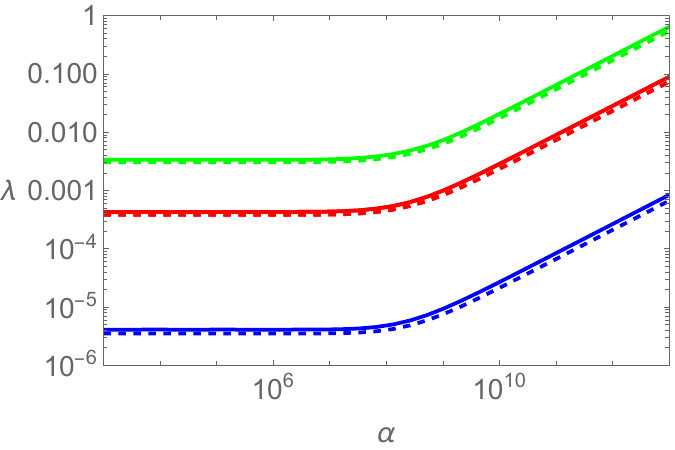}}
    \caption{$r$ vs. $n_s$ (a), $r$ vs. alpha (b), $\alpha$ vs. $n_s$ (c), $\lambda$ vs. $\alpha$ (d) for the PV potential with $k=2$ (blue), $k=4$ (red), $k=8$ (green) for $N_e =60$ (thick)  and $N_e = 70$ (dashed). The dots represents the predictions of the of the original PV potential in the same color code. For each value of $k$ the plot is obtained fixing $\lambda$ by imposing the condition on the observed amplitude of the scalar perturbations $A_s = 2.1 \cdot 10^{-9}$ and varying $\alpha$ in the range $0<\alpha<10^{13}$ (i.e. from small to large couplings of the higher order term). The gray regions indicate the 95\% (dark-gray) and 68\% (light-gray) confidence levels (CL), respectively, based on the latest combination of Planck, BICEP/Keck, and BAO data \cite{BICEP:2021xfz}.}
    \label{fig:slow_roll_peebles}
\end{figure*}

We plot the original PV potential (black), and the modified PV potential in eq. \eqref{eq:fractional_potential} (blue) with $\alpha =2\cdot 10^8$, $q=k=4$, predicting viable CMB observables  $r = 0.017$, $n_s=0.966$ (at $N_e = 60$) and a quintessential tail that solves the coincidence problem with the mass scale $M=1.38\cdot 10^{-13}$. We also show $\phi_N$ (star) and $\phi_{end}$ (dot) in the same color code. $\phi_N$ is not visible for the original PV potential as it lies at $V(\phi_N)\sim 10^{-8}$. 
The qualitative shape of the potential does not depend on the specific parameters.
The value $\phi_{end}$ is such that $\epsilon(\phi_{end})=1$ which corresponds to the end of slow-roll inflation, while $\phi_N$ is fixed by imposing $N_e \sim 60-70$, i.e. the number of $e$-folds between the time at which the CMB pivot scale $k=0.05$ Mpc leaves the horizon, and the end of inflation.

We notice that, while the potential is modified at the inflation scale, it remains unchanged in the tail. This is indeed expected. By taking the limit $\phi\rightarrow-\infty$ of eqs. \eqref{eq:fractional_potential} and \eqref{eq:peebles},  we see that $U(\phi)\rightarrow \frac{1}{8\alpha}$ which sets the height of the inflationary plateau. On the other hand, for $\phi \rightarrow \infty$ we get $U(\phi) \sim V(\phi)$, which reproduces the behavior of the original potential.

We now study the details of the inflationary phenomenology.

We compute the CMB observables in the slow-roll approximation by means of the slow roll parameters:
\begin{align}
\epsilon(\phi) &\equiv \frac{1}{2}\qty(\frac{U'(\phi)}{U(\phi)})^2, \\
\eta &\equiv \frac{U''(\phi)}{U(\phi)}.
\end{align}
In this approximation we can compute the tensor-to-scalar ratio $r$, the spectral index $n_s$, the amplitude of the scalar perturbations $A_s$ and the number of $e$-folds $N_e$ as:
\begin{align}
  r &= 16\epsilon (\phi_N)  \\
  n_s &= 1-2\eta(\phi_N)+6\epsilon(\phi_N)\\
  A_s &= \frac{U(\phi_N)}{24\pi^2 \epsilon(\phi_N)}\\
  N_e &= \int_{\phi_{end}}^{\phi_{N}}d\phi \frac{U(\phi)}{U'(\phi)} 
\end{align}
For our choice of the Einstein frame potential, we obtain:
\begin{align}
\label{eq:epsilon_peebles}
\epsilon(\phi_N) &=
 \frac{k^2 \phi_N ^{2 k-2}}{2 \left(M^k+\phi_N ^k\right)^2 \left(8 \alpha  \lambda ^k \phi_N ^k+8 \alpha  \lambda ^k M^k+1\right)^2} \, \\
\label{eq:eta_peebles}
\eta(\phi_N) &=
      \frac{k \phi_N ^{k-2} }{\left(M^k+\phi_N ^k\right) \left(8 \alpha  \lambda ^k \phi_N ^k+8 \alpha  \lambda ^k M^k+1\right)^2} \times \nonumber\\
      & \times \left(-8 \alpha  k \lambda ^k \phi_N ^k-8 \alpha  \lambda ^k \phi_N ^k+8 \alpha  k \lambda ^k M^k \right. \nonumber\\
      & \qquad \left. -8 \alpha  \lambda ^k M^k+k-1\right)
\end{align}
\begin{align}
    N_e &=  \left[\frac{\phi ^{2-k}}{2 k} \left(\frac{\phi ^k \left(16 \alpha  \lambda ^k
   \phi ^k+k+2\right)}{k+2}-\frac{16 \alpha  \lambda ^k M^{2
   k}}{k-2}+ \right. \right. \nonumber\\
   & \qquad \left. \left. 2 M^k \left(8 \alpha  \lambda ^k \phi
   ^k+\frac{1}{2-k}\right) \right) \right]^{\phi_N}_{\phi_{end}}.
\end{align}
In the strong coupling limit for $\alpha \to \infty$, $\phi_N \to -\infty$, and neglecting the contribution of $\phi_{end}$ the inflationary observables can be approximated as: 
\begin{align}
N_e&\simeq\frac{8\alpha\lambda^k{\phi_N}^{k+2}}{k(k+2)},\\
r& \simeq  \frac{1}{12 \pi ^2 \alpha  A_s}, \label{eq:r:limit:1}
\\
n_s&\simeq1-\frac{k+1}{k+2}\frac{2}{N_e}, \label{eq:ns:limit:1} \\
A_s&\simeq\frac{k+2}{12\pi^2 k}\lambda^k N_e \qty(\frac{k(k+2)N_e}{8\alpha\lambda^k})^\frac{k}{k+2}. 
\end{align}
We plot in Fig.~\ref{fig:slow_roll_peebles} the numerical results for slow-roll inflation using this model. Notice that the number of e-folds is chosen in the range $N_e =60-70$ instead of the usual $N_e =50-60$, due to kination contribution during the Universe expansion, which is absent in models that consider an oscillatory reheating mechanism. The extra contribution to the number of e-folds can be computed as in  ref.~\cite{Drewes:2017fmn}: 
\begin{equation}
\Delta N = \frac{1-3w}{12(1+w)}\ln{\qty(\frac{\rho_{reh}}{\rho_{end}})} \simeq \frac{1}{3}\ln{\qty(\frac{U_{end}^{1/4}}{T_{reh}})},
\end{equation}
where $\rho_{reh}$ and $\rho_{end}$ are respectively the energy density of the universe at the time of reheating and the end of inflation, $U_{end}$ is the potential energy density at the end of inflation and $T_{reh}$ the reheating temperature. In the last expression, we used the fact that $w\equiv P/\rho=1$ during kination. Notice, moreover, that $\Delta N$ is zero in standard Big Bang cosmology because $w=1/3$ during radiation domination.
For kination, the above gives \mbox{$\Delta N \sim 10$} for $U_{end}^{1/4} \sim 10^{16}$ GeV and $T_{reh}\sim 10^3 $ GeV, which justifies the choice $N_e \sim 60-70$. The relation between $\lambda, \alpha$ is imposed by fixing $A_s = 2.1 \cdot 10^{-9}$ \cite{Planck2018:inflation}. From Fig. \ref{fig:slow_roll_peebles} we can see that agreement with data can easily be achieved with $\alpha \gtrsim 10^8$. To conclude, we note that the asymptotic results given in eqs. \eqref{eq:r:limit:1} and \eqref{eq:ns:limit:1} are exactly the same as the ones obtained in \cite{Dioguardi:2023} because in the inflationary region the models are equivalent in the large $\alpha$ region. 

\subsection{Kination}\label{sec:kination}
\begin{figure*}[t]
    \centering
    
    \subfloat[]{\includegraphics[width=0.4\textwidth]{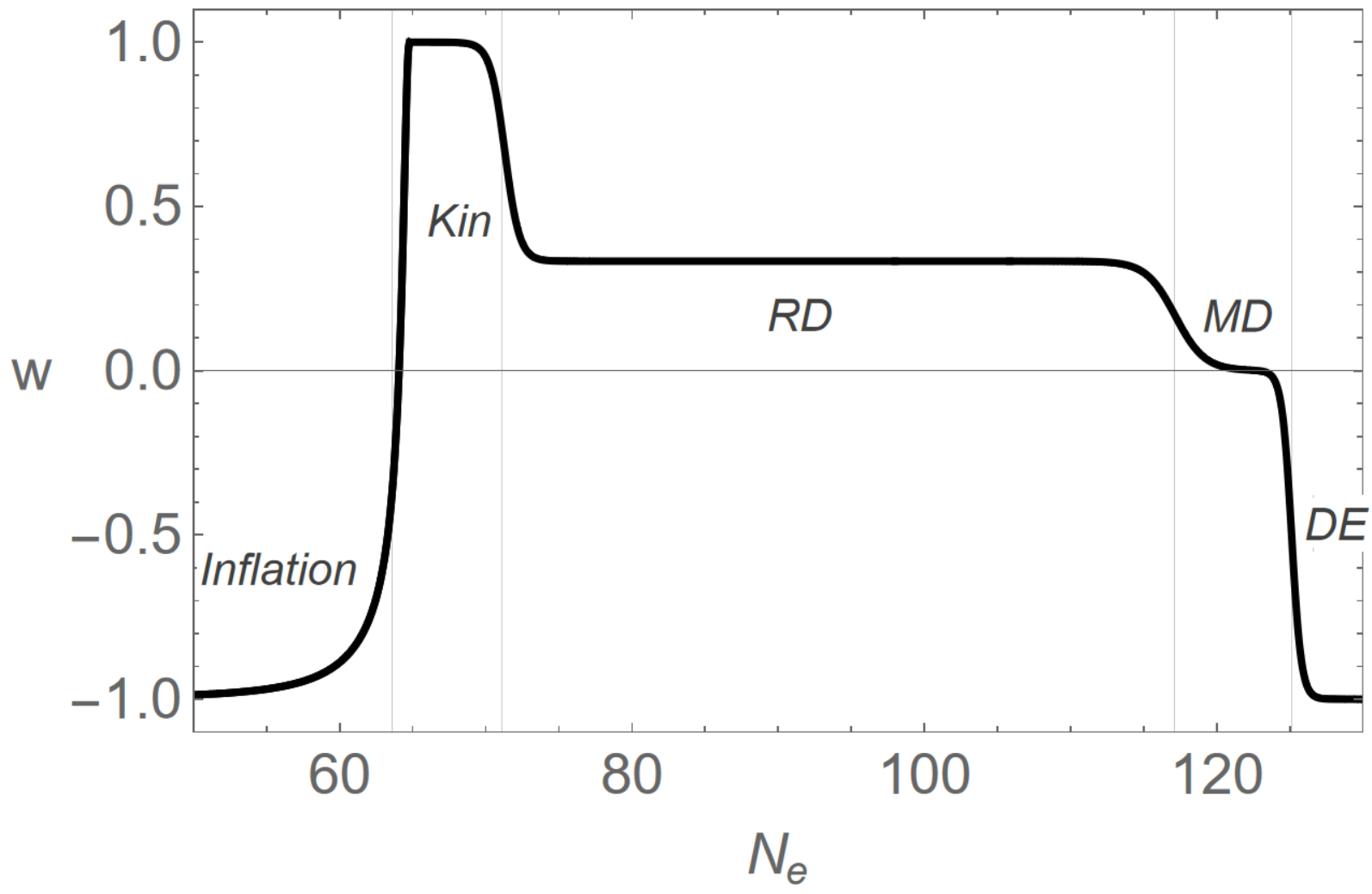}}%
    \qquad
    \subfloat[]{\includegraphics[width=0.418\textwidth]{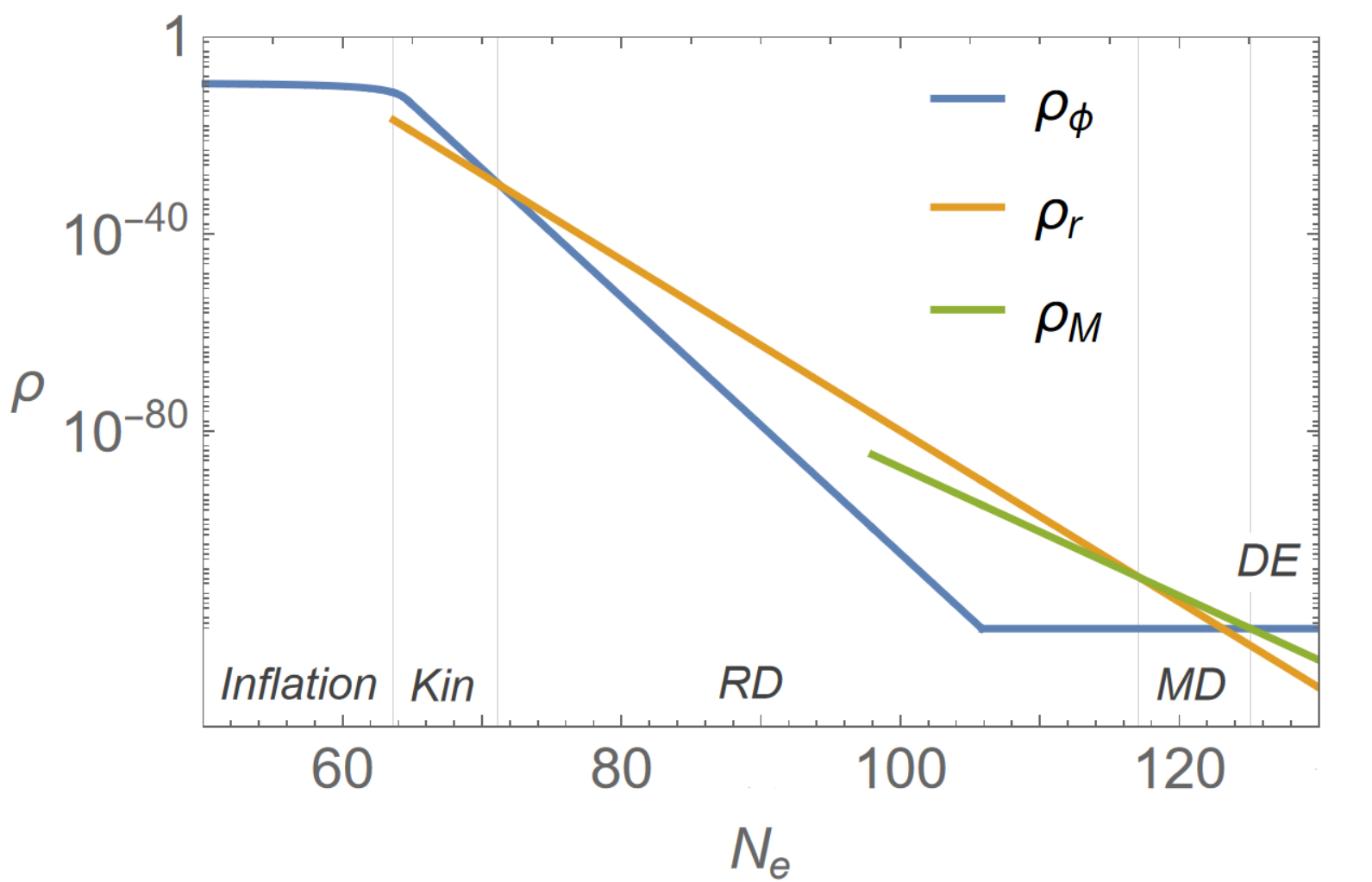}}%
    \caption{a) Evolution of the barotropic parameter of the universe in terms of the elapsing $e$-folds number $N_e$ for the benchmark potential in Fig.\ref{fig:U_peebles_potential} with $\alpha = 2\cdot 10^8$, $q=k=4$ and $M = 1.38\cdot 10^{-13}$. The plot shows the natural appearance of a kination phase $w=1$ right after the end of inflation (Kin). As radiation domination (RD) begins $w$ drops to $1/3$. After that we have matter domination (MD) with $w=0$. Finally in the recent time the scalar field energy density becomes again the dominant source of energy-density in the universe (DE). Today we have $w =-\Omega_\Lambda \simeq -0.7$. The vertical lines denote the corresponding $e$-folds number $N_e$ for transitions between different epochs. b) Evolution of the energy densities for the scalar field $\rho_\phi$, the radiation fluid $\rho_r$ and the matter fluid $\rho_M$ in units of $m_P^4=1$.  The vertical lines denote the corresponding $e$-folds number $N_e$ for the transitions between different epochs. It is assumed that radiation is originally generated at the end of inflation (e.g. by Ricci reheating \cite{Dimopoulos:2018wfg,Opferkuch:2019zbd,Bettoni:2021zhq}).}
    \label{fig:w_evo}
\end{figure*}

Right after the end of inflation the equation of motion for the Einstein frame scalar field:
\be
\Ddot{\phi}+3H\dot{\phi}+V'(\phi)=0 \, ,
\ee
becomes dominated by kinetic energy and can be approximated as:
\be \label{eq:friedmann_kin}
\Ddot{\phi}+3H\dot\phi \simeq 0 \, .
\ee 
It is evident that the evolution of the scalar field is now oblivious of the potential, and hence it is in essence model independent. The behaviour of the rolling field is therefore generic.

The solution of the above for $\dot \phi$ is given by 
\be\label{eq:kinetic_term}
\dot\phi=\sqrt{\frac{2}{3}}\,\frac{1}{t} \, .
\ee
Hence, by assuming that the scalar field behaves as a perfect fluid we have:
\be
\rho_\phi = \frac12\dot\phi^2+V(\phi)\sim \frac12\dot \phi^2 \propto a^{-6} \, ,
\ee
with $a \propto t^{1/3}$, the scale factor of the universe during kination. Since from the Friedmann equations we have that $\rho \propto a^{-3(1+w)}$, the scalar field in this epoch behaves as a barotropic fluid with $w=1$.
By direct integration we have from eq.~\eqref{eq:kinetic_term}:
\be \label{eq:scalar_kination}
\phi(t) = \phi_{end} + \sqrt{\frac{2}{3}}\ln{\qty(\frac{t}{t_{end}})}.
\ee

We show in Fig.~\ref{fig:w_evo}(a) the evolution of the of the barotropic parameter in terms of the $e$-folds number $N_e$ for the benchmark potential in Fig.~\ref{fig:U_peebles_potential} with $\alpha = 2\cdot 10^8$, $q=k=4$ and $M = 1.38\cdot 10^{-13}$. The plot shows the natural appearance of a kination phase $w=1$ right after the end of inflation.

Kination finally needs to end in order to let the Hot Big Bang (HBB) take place. This has to happen before Big Bang Nucleosynthesis (BBN), so to reproduce the correct abundances of primordial nuclei.
Since the radiation energy density scales as $\rho_r \propto a^{-4}$, as the Universe expands radiation will eventually become the dominant source of the energy-matter content of the Universe. This process is called reheating, and it can happen right after inflation ends 
or later on depending on the energy density of radiation at the end of inflation. The main constraint is that transition from kination to radiation domination has to happen before BBN takes place. 

\subsection{Reheating}\label{sec:reheating}
We consider that radiation first appears at the end of inflation.
We define:
\be \label{eq:density_radiation_grav}
\Omega^{grav}_r|_{end}  \equiv \frac{\rho_{r}^{grav}}{\rho}|_{end}
\ee
the density parameter for radiation at the end of inflation. Its specific value will depend on the choice of the reheating mechanism, and in particular the following general condition holds:
\be \label{eq:density_param_grav}
\Omega^{grav}_r|_{end}\leq\Omega_r|_{end}\leq 1\,,
\ee
where the upper bound is found by choosing prompt reheating, while the lower bound is set by gravitational reheating which is the weakest reheating mechanism and hence produces the lowest possible value for the density parameter of radiation.
The HBB starts with reheating and takes place at time given by:
\be
t_{reh} = \frac{t_{end}}{(\Omega_r|_{end})^{3/2}}\,,
\ee
which implies:
\be \label{eq:scalar_reheating}
\phi_{reh} = \phi_{end}-\sqrt{\frac{3}{2}}\ln{(\Omega_r|_{end}})\,.
\ee
Note that $\phi_{reh}$ only depends on the potential $U(\phi)$ implicitly through the end of inflation, but its derivation is model-independent.
Now if we assume that radiation has the time to thermalize by the beginning of radiation domination then we can compute the reheating temperature as:
\be\label{eq:T_reheating}
T_{reh} =\qty(\frac{30}{\pi^2 g_*^{reh}}(\Omega^{end}_r)^3 \rho_\phi^{end})^{1/4},
\ee
where $g_*^{reh}$ are the effective relativistic degrees of freedom at the time of reheating. In the following we keep $\Omega_r^{end}$ as a free parameter, i.e. we do not assume any specific reheating mechanism and proceed to find a lower bound for the reheating temperature from the constraints on overproduction of gravitational waves during the kination phase, as explained in appendix~\ref{sec:appendix}.
It is, in fact, well-known that a kination phase induces a peak in the amplitude of the gravitational wave background \cite{HaroCases:2020nsn}. This peak has to be constrained in order to do not spoil the observations on Big Bang Nucleosynthesis. If detected in the future this would represent a confirmation of physics beyond $\Lambda$CDM and an hint for the viability of quintessential inflation models.

\subsection{Quintessence}\label{sec:quintessence}

During radiation domination as the Universe keeps expanding; the field loses its kinetic energy while rolling down the field until it freezes. Since the potential is still negligible the EoM for the scalar field is still given by eq.~\eqref{eq:friedmann_kin}, but $H$ is now determined by the radiation content of the Universe so the equation yields:
\be
\dot\phi = \sqrt{\frac{2}{3}\frac{t_{reh}}{t^3}}\,.
\ee
Hence, by integrating it we get:
\be \label{eq:scalar_radiation}
\phi(t) = \phi_{reh}+2\sqrt{\frac{2}{3}}\qty(1-\frac{t_{reh}}{t})\,.
\ee
By using eqs.~\eqref{eq:scalar_reheating} and \eqref{eq:scalar_radiation}, we get that the scalar field freezes at $t\gg t_{reh}$ at the value:
\be \label{eq:frozen_scalar}
\phi_F = \phi_{end} +\sqrt{\frac{2}{3}}\qty(2-\frac{3}{2}\ln\Omega_r^{end})\,.
\ee
In order to have a working quintessential inflation mechanism, we need to satisfy two more requirements. The first one is that the value of the scalar field potential energy-density at freezing must match the value of the observed energy density of dark energy today (coincidence requirement), that is:
\be \label{eq:coincidence}
U (\phi_F) =\lambda^k\frac{M^{k+q}}{\phi_F^q+M^q} = \rho_0 \sim 7 \cdot 10^{-121} . 
\ee
Second, we need to check that the barotropic parameter at  present $w_0$ for the quintessential tail is within the Planck observational bounds \cite{Planck:cosmo}. 
In order for this to happen we need to make sure that the scalar field unfreezes only at the time of matter-dark energy equality. In other words, the field stays frozen until recent times and only unfreezes when it becomes dominating.

It is a well-known fact that in quintessence models, a competitor attractor behavior can appear. This attractor behavior is referred to in the literature as a tracker, if it leads to eventual quintessence domination.
The condition for the tracker to appear is given by:
\be
\Gamma \equiv \frac{U U''}{(U')^2} > 1.
\ee
Note that, for an inverse-power-law potential like the one we have chosen, the tracker condition gives $\Gamma=\frac{q+1}{q}$ which  implies that the tracker is always present. For the case $q=4$ the energy density of the tracker solution scales as:
\begin{equation}\label{eq:tracker_solution}
\rho_T \sim \qty(\frac{M^8} {\alpha t^4})^{1/3}, 
\end{equation}
which scales slower than a matter-dominated background.

While the field is frozen at $\phi_F$ it has a constant contribution to the energy-density. However, if it hits the tracker before becoming dominant it unfreezes and starts following the tracker solution. Eventually,the scalar field becomes dominant with a barotropic parameter $w = -\frac13$, which is not acceptable given the observational bound $w_{DE}<-0.95$ \cite{Planck:cosmo}. If, on the other hand, the value of $U(\phi_F)$ is small enough i.e. 
\begin{equation}\label{eq:tracker_condition}
 U(\phi_F (t_{eq}))<\rho_T(t_{eq}), 
\end{equation}
then the scalar field unfreezes only at the matter-dark energy equality. 
After dominating, the field thaws and starts slow-rolling down its potential. The slow-rolling quintessence evolves as a fluid with a barotropic parameter given by:
\be \label{eq:barotropic_phi}
w_\phi = \frac{\frac{1}{2}\dot\phi^2 - U(\phi)}{\frac{1}{2} \dot \phi^2 + U(\phi)} \simeq -1\, , 
\ee
since the field changes very slowly with time. Note that, while close to $-1$, the barotropic parameter would be slightly larger, which, if confirmed by observations, would distinguish the model from $\Lambda$CDM.\footnote{Indeed, recent DESI observations seem to suggest that quintessence is favoured over $\Lambda$CDM \cite{DESI:2024mwx}.}

Imposing \eqref{eq:tracker_condition} is equivalent to impose a lower bound on the mass scale $M=M_{min}$ in our potential (see Table \ref{tab:peebles_k4}).
It can be checked that field does not evolve substantially from matter-dark energy equality to the present day. Therefore, it is safe to consider $\phi_0 \simeq \phi_F$ and $w_{\phi} \approx -1$.
\begin{table}\centering
   \begin{tabular}{|p{0.7cm}||p{1.7cm}|p{0.8cm}|p{1.7cm}|p{1.5cm}|}
 \hline
 $\alpha$ & $M_{min}[GeV]$ & $\phi^{min}_F$ & $T_{reh}^{max}[GeV]$ & $T^*_{reh}[GeV]$\\ 
 \hline
  $10^8$         & $2.0\cdot 10^5$ & 4.01 & $4.6\cdot 10^2$ & $1.3\cdot 10^7$ \\ 
  $10^{12}$      & $2.7\cdot 10^4$ & 4.01 & $2.3\cdot 10^3$ & $1.9\cdot 10^5$\\ 
  $10^{16}$       & $2.7\cdot 10^3$ & 4.01 & 0.27           & $5.9\cdot 10^2$\\
 \hline
\end{tabular} 
\caption{Results for reheating for the PV potential \eqref{eq:peebles} with $q=k=4$. The estimated maximum reheating temperature for the model $T^{max}_{reh}$ is below the lower bound $T^{*}_{reh}$ for every choice of $\alpha$, this implies that in general the model cannot account for the BBN constraints on overproduction of GWs (see appendix \ref{sec:appendix}). However compatibility with observations is restored if one assumes production of heavy particles in the early universe \cite{saviour}.}
\label{tab:peebles_k4}
\end{table}
We show in Table \ref{tab:peebles_k4} the results for the case $k=q=4$. We only report the values of $\alpha$ that allow good predictions for the CMB observables. The value $M_\text{min}$ is the minimum value of the mass scale for the potential in eq.~\eqref{eq:peebles} necessary to avoid the tracker solution and let the field unfreeze only at matter-dark energy equality. The corresponding $\phi_F^\text{min}$ is obtained by imposing the solution of the coincidence problem in eq.~\eqref{eq:coincidence}. The reheating temperature $T_{reh}^{max}$ is computed by means of eq.~\eqref{eq:T_reheating} by setting $M=M_\text{min}$ and $\phi=\phi_\text{min}$.

Finally, we need to check if the model can satisfy the constraints on $T_{reh}$ coming from gravitational-wave (GW) overproduction. During kination, the $w = 1$ stiff period induces a spike in the density of the GWs, potentially spoiling BBN. This can be avoided if $T_{reh}$ is large enough. The details on the calculation can be found in appendix \ref{sec:appendix} (see eq. \eqref{eq:GW_integral}-\eqref{eq:Omega_P}) , while we show in Table \ref{tab:peebles_k4} an estimate of the lower bound $T^*_{reh}$ (depending on the value of $\alpha$) that avoids GW overproduction. 

Unfortunately, we always have $T_{reh}^{max}<T^*_{reh}$ hence the model is not viable in general. However, compatibility with observation can be restored if we assume the gravitational production of very massive particles after the inflationary era \cite{saviour}. How much the bound on reheating temperature relaxes,  depends on the exact properties of such heavy particles, like mass, heating efficiency etc. However, regardless of the aforementioned details, an indicative validity range can be estimated to be \cite{saviour} 
\be
\text{1 MeV} \leq T_{reh} \leq 5\cdot 10^7 \text{GeV} \, , \label{eq:saviour}
\ee
which makes the results computed in Table \ref{tab:peebles_k4} viable.

\subsection{The case of the exponential tail}\label{sec:exponential}

\begin{figure}[t]
    {\includegraphics[width=0.45\textwidth]{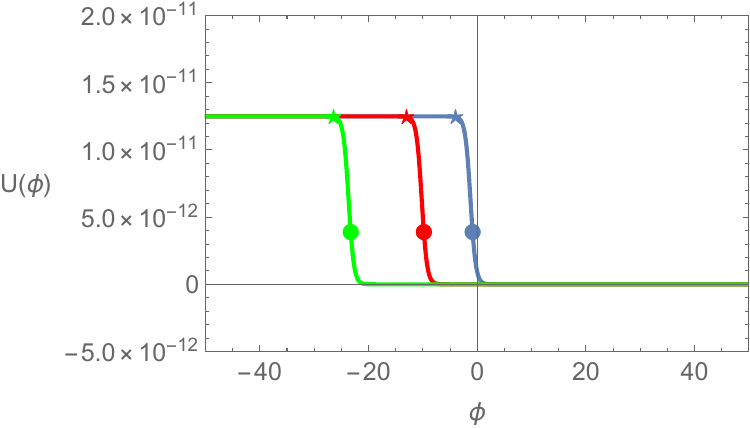}}%

    \caption{Exponential potential for $M=10^{-3}$ (blue), $M=10^{-5}$ (red), $M=10^{-8}$ (green) and $\alpha = 10^{10}$ and corresponding $\lambda =2.05$ fixed by setting $A_s \sim 2.1 \cdot 10^{-9}$. We also show $\phi_N$ (star) and $\phi_{end}$ (dot) in the same color code. Changing $M$ amounts to a shift in the potential along $\phi$-axis.} 
    \label{fig:U_exponential_potential}
\end{figure}

Most popular quintessence models assumes an exponential tail in the form:
\be
V(\phi) = M^4 e^{-\lambda\phi}. \label{eq:V:exp}
\ee
By choosing a Jordan frame potential as in eq. \eqref{eq:V:exp} we get the Einstein frame potential
\begin{figure*}[t]
    \centering
    
    \subfloat[]{\includegraphics[width=0.4\textwidth]{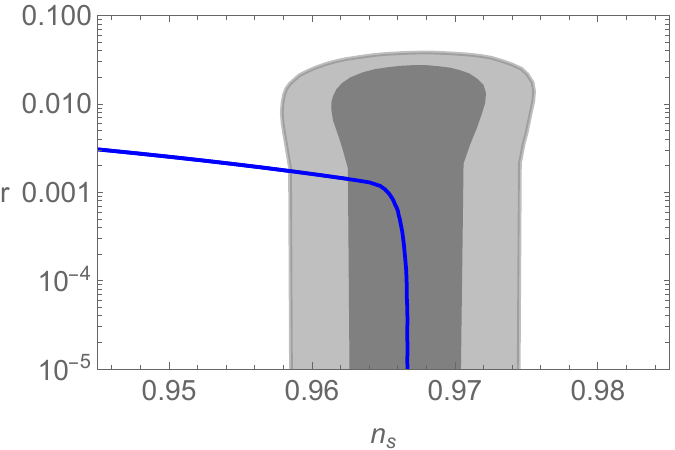}}%
    \qquad
    \subfloat[]{\includegraphics[width=0.4\textwidth]{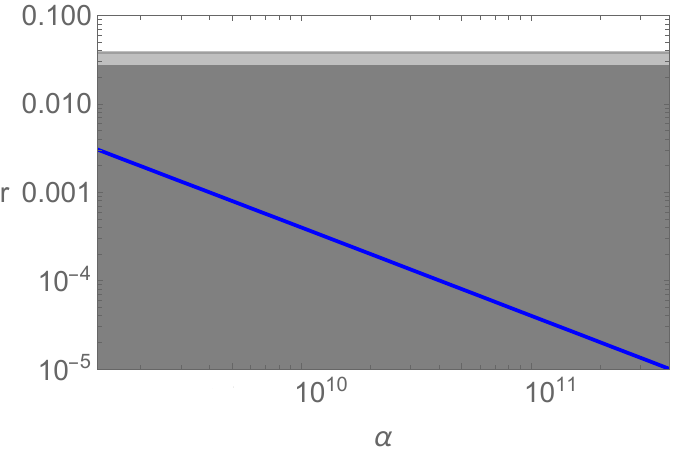}}%
    
    \subfloat[]{\includegraphics[width=0.4\textwidth]{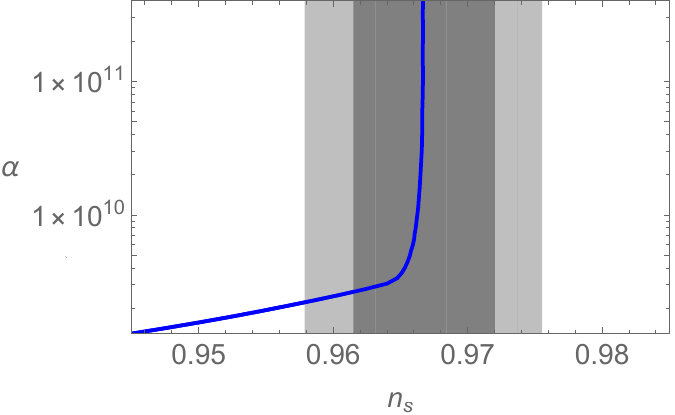}}%
    \qquad
    \subfloat[]{\includegraphics[width=0.36\textwidth]{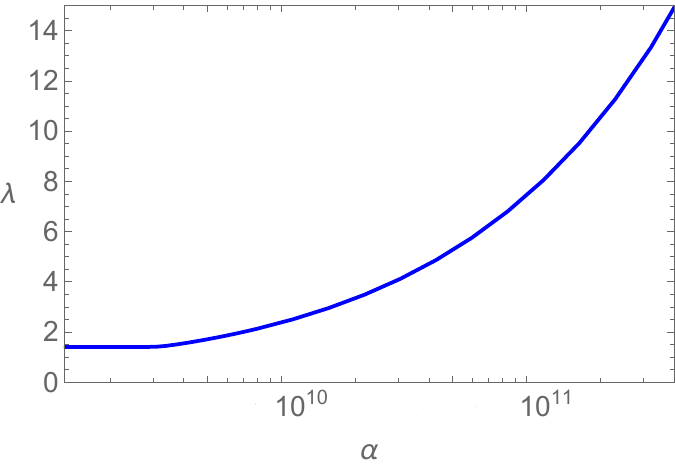}}
    \caption{$r$ vs. $n_s$ (a), $r$ vs. $\alpha$ (b), $\alpha$ vs. $n_s$ (c), $\lambda$ vs. $\alpha$ (d) for the quadratic model with $V(\phi) = V_0 e^{-\lambda \phi}$ for $N_e =60$. The relation between $\alpha$ and $\lambda$ is imposed by fixing $A_s \sim 2.1\cdot 10^{-9}$. The gray regions indicate the 95\% (dark-gray) and 68\% (light-gray) confidence levels (CL), respectively, based on the latest combination of Planck, BICEP/Keck, and BAO data \cite{BICEP:2021xfz}.}
    \label{fig:slow_roll_exp}
\end{figure*}
\be \label{eq:einstein_potential_exp}
U(\phi)=\frac{M^4 e^{-\lambda  \phi }}{8 \alpha M^4 e^{-\lambda  \phi }+ 1},
\ee
and the corresponding the slow-roll parameters
\begin{align}\label{eq:epsilon_exp}
\epsilon(\phi)&=\frac{\lambda ^2}{2} \qty(\frac{1}{1+8\alpha M^4 e^{-\lambda \phi}})^2, \\
\eta(\phi)&=\frac{\lambda ^2 e^{\lambda  \phi } \left(1-8 M^4 \text{$\alpha $}e^{-\lambda  \phi }\right)}{\left(8 M^4 \text{$\alpha $}e^{-\lambda  \phi }+1\right)^2}.
\end{align}
From eq.~\eqref{eq:epsilon_exp} we see that:
\begin{align}
 \epsilon(\phi) &\sim 0, & \text{for}\;&\hspace{3pt} \phi\to -\infty\\
 \epsilon(\phi) &\sim \frac{\lambda^2}{2} & \text{for}\;&\hspace{3pt} \phi\to \infty\,,
\end{align}
which implies that to end slow-roll during inflation and have a graceful exit we need $\lambda>\sqrt{2}$.
While inflation works very well for this choice of parameters, as shown in Fig.~\ref{fig:slow_roll_exp}, it does not provide a good tail for quintessence. The reason can be seen from Fig.~\ref{fig:U_exponential_potential}. 
The plot shows the change in the potential by varying the parameter $M$. The net result of changing $M$ amounts to a translation along the $\phi-$axis, while $\phi_{end} -\phi_{N}$ remains constant (after fixing the other parameters $\alpha$,$\lambda$,$N_e$), yielding the same results for the inflationary observables. The immediate consequence is that $M$ cannot be used as a parameter to fix, in order to match the observed cosmological constant. 

To see this explicitly, consider the tail
of eq.~\eqref{eq:einstein_potential_exp}. For $\phi\rightarrow +\infty$, we have:
\be
U(\phi) \sim M^4 e^{-\lambda \phi}.
\ee
Once again we want to understand if we can solve the coincidence problem:
\be
U(\phi_F) = M^4 e^{-\lambda\phi_F} = \rho_0 \sim 7\cdot 10^{-121},
\ee
where $\phi_F$ is still given by eq.~\eqref{eq:frozen_scalar}. 
For the sake of simplicity, we now consider the case that maximizes eq.~\eqref{eq:frozen_scalar}, i.e. gravitational reheating.
It can be computed that:
\be
\phi_F \sim \phi_{end} + 40,
\ee
where $\phi_{end}$ is obtained by solving $\epsilon(\phi)=1$.
Hence, we get:
\be\label{eq:coincidence_exp}
U(\phi_F) = \frac{\sqrt{2}\lambda -2}{16\alpha}e^{-\lambda 40}.
\ee
This implies that the only parameter in the model is $\lambda$ and the two scales (inflation and dark-energy) cannot be decoupled.
Even in the maximal case we cannot solve the coincidence problem unless $\lambda\sim 6.3$ which is too large and cannot reproduce dark-energy behavior.
It is then impossible to use this model to achieve a viable quintessential inflation scenario.

\section{\boldmath
\textbf{Quintessential inflation for $F(R_X)_{>2}$}}\label{sec:peebles_log}

In this section, we discuss for completeness quintessential inflation for the higher-order case $F(R,X)_{>2}$. We carry the same analysis of section~\ref{sec:peebles} by computing the inflation and dark-energy observables and consider the bounds from overproduction of gravitational waves, in order to constrain the parameter space and prove the viability of quintessential inflation for this class as well.

For this class of functions, the behavior of  $G(\zeta)$ changes drastically and one can see that the equation $G(\zeta)=0$ always admits a solution for some $\zeta_0 > 0$ (see \eqref{eq:G=V}). For any $\zeta > \zeta_0$, the $G(\zeta)$ function is negative. As shown in \cite{Dioguardi:2023}, this configuration works particularly well when considering negative and unbounded from below Jordan frame potentials $V(\phi)$. In fact, such a choice generates an Einstein frame potential $U$ which is positive definite and has an asymptotically flat plateau that allows to perform inflation for $\zeta \rightarrow \zeta_0$ (i.e. $V(\phi) \to 0$). At the same time for $\zeta\rightarrow +\infty$, the potential $U$ exhibits a tail that asymptotically approaches zero, giving the opportunity to mimic the quintessence behavior without the necessity of introducing a decaying tail behavior to begin with.

In order to give a concrete example, we now focus on the case:
\be
F(\zeta)= \zeta +\alpha \zeta^2 \ln(\beta\zeta),
\ee
with $\beta>\alpha/e$,
in order to ensure that $F'>0$ for any $\zeta>\zeta_0$.
This specific example allows for an exact solution for eq. \eqref{eq:G=V}, which reads:
\be
  \zeta -\alpha \zeta^2 = V(\phi).
\ee
The above equation admits two solutions:
\be
\zeta_\pm=\frac{1 \pm \sqrt{1 - 4 \alpha  V(\phi)}}{2 \alpha } = \frac{1 \pm \sqrt{1 + 4 \alpha  |V(\phi)|}}{2 \alpha },
\ee
but the consistency constraints $\zeta \geq \zeta_0 = 1/\alpha$ and $V(\phi)<0$, suggest 
\be
 \zeta = \zeta_+ = \frac{1 + \sqrt{1 + 4 \alpha  |V(\phi)|}}{2 \alpha },
\ee
is the only available solution. Using such a solution, we can provide the exact expression of the Einstein frame scalar potential:
\bea
 U(\phi) &=& \frac{| V(\phi )| }{2} \Bigg[4 \alpha  | V(\phi )|  \ln
   \left(\frac{\beta  \qty(1+\sqrt{1+ 4 \alpha  | V(\phi )|
   })}{2 \alpha }\right)  \nonumber\\
&& \phantom{\frac{| V(\phi )| }{2} \Bigg[}  +2 \alpha  | V(\phi )|
   +\sqrt{4 \alpha  | V(\phi )| +1}-1 \Bigg]^{-1} \label{eq:einstein_potential_log}
\eea
For $V(\phi) \to 0$, the above 
behaves as
\be
U(\phi) \approx \frac{1}{8 \alpha  \left(\ln \left(\frac{\beta }{\alpha}\right)+1\right)} \left[1 + \frac{ \alpha   V(\phi)}{2  \left(\ln \left(\frac{\beta }{\alpha   }\right)+1\right)} \right],
\ee
while for $V(\phi) \to -\infty$, as
\be\label{eq:tail_log}
U(\phi) \approx \frac{1}{8 \alpha  \ln \left(\beta  \sqrt{\frac{|V(\phi)|}{\alpha
   }}\right)} = \frac{1}{8 \alpha \left( \frac{1}{2} \ln \left(|V(\phi)| \right) + \ln \left( \frac{\beta }{\sqrt\alpha} \right) \right) },
\ee
where we emphasize once again that $V(\phi) <0$.
We conclude this preliminary discussion by noting that, if we plug-in an exponential potential with the an asymptotic behaviour
\be
V(\phi) \approx -e^{f(\phi)}, \label{eq:V:exp:approx}
\ee
into eq. \eqref{eq:tail_log}, then the tail of the potential behaves as 
\be
U(\phi) \approx \frac{1}{f(\phi)}. \label{eq:generic:tail:F>2}
\ee
We will use this last result as a guide to construct a working model of quintessential inflation, as we will see in the following subsection.

\subsection{Exponential PV potential}
Given the result of eq. \eqref{eq:generic:tail:F>2}, we choose an exponential version\footnote{We added $1$ to the definition of the potential in eq. \eqref{eq:V:exp:approx} in order to ensure that $V(\phi) \to 0 $ for $\phi \to -\infty$.} of the Peebles-Vilenkin ($e$PV) potential used in section \ref{sec:peebles}
\be\label{eq:peebles:exp}
V(\phi) = 1 - e^{V_{PV}(\phi)} \, ,
\ee
with
\be
V_{PV}(\phi) =
\begin{cases} 
      \frac{\mu^k}{M^q (\phi^k + M^k)} & \phi\leq 0, \\
       \quad & \quad \\
      \frac{\phi^k + \mu^k}{M^{k+q}} & \phi > 0\,. \\
 \end{cases}
\ee
It is possible to show numerically that the parameter $\beta$ appearing in eq.~\eqref{eq:einstein_potential_log} has a negligible effect on the computation of the CMB observables. Moreover, one can see from eq.~\eqref{eq:tail_log} that for realistic choices of $\alpha,\beta$ the dominant term in the denominator will be given by the $\ln (|V|)$ contribution. Thus, in what follows, we can safely set $\beta = \alpha$ without loss of generality.
With this setup it is possible to produce viable predictions for the CMB observables by fixing the two mass scales $M,\mu$ without requiring extreme fine tuning to solve the coincidence problem.
\begin{table}\centering
   \begin{tabular}{|p{1.7cm}||p{1.7cm}|p{1.7cm}|p{1.7cm}|}
 \hline
 $\alpha$ & $r$ & $n_s$ & $\mu / M $ \\ 
 \hline
  $10^8$  & $0.03$              & $0.963$ & 0.06 \\
  $10^{12}$ & $4\cdot 10^{-6}$  & 0.972   & $2.5\cdot 10^{-5}$ \\
  $10^{16}$ & $4\cdot 10^{-10}$ & 0.972   &  $2.5\cdot 10^{-9}$\\

\hline
\end{tabular} 
\caption{Results for inflation for the $e$PV potential in eq.~\eqref{eq:peebles:exp} with $k=q=4$ at $N_e = 60$. Predictions for inflation lay inside the $2\sigma$ region for $r,n_s$.  The ratio $\mu/M$ is fixed by requiring $A_s =2.1\cdot 10^{-9}$.}
\label{tab:peebles_exp_k8_inflation}
\end{table}
However, as we can see from Table \ref{tab:peebles_exp_k8_inflation}, this setup is physical only for very large $\alpha$.
For this reason, we only compute the slow-roll parameters in the $\alpha \to \infty$ regime:
\bea
\epsilon(\phi_N) \sim \frac{1}{8} \alpha ^2 k^2 \mu ^{2 k} \phi ^{-2 (k+1)} M^{-2 q} ,\\
\eta(\phi_N) \sim -\frac{1}{2} \alpha  k (k+1) \mu ^k \phi ^{-k-2} M^{-q} \,,
\eea
and the number of e-folds can be integrated explicitly giving: 
%
\be
N_e \sim \frac{2 \mu ^{-k} \phi ^{k+2} M^q}{\alpha  k (k+2)}.
\ee
Finally, the CMB observables read:
\bea
r &\sim&  \frac{1}{12 \pi ^2 \alpha  A_s}, 
\\
n_s &\sim& 1- \frac{k+1}{k+2}\frac{2}{N_e},\\
A_s &\sim & \frac{2^{-\frac{5 k+8}{k+2}} \mu ^{-2 k} M^{2 q} \left(\alpha  k (k+2) N_e \mu ^k M^{-q}\right){}^{\frac{2 (k+1)}{k+2}}}{3 \pi ^2 \alpha ^3 k^2}.
\eea
We note that the limits for $r$ and $n_s$ are the same as the ones shown in eqs. \eqref{eq:r:limit:1} and \eqref{eq:ns:limit:1}. Indeed, it can be easily proven that the current setup and the one in section \ref{sec:inflation} are actually equivalent in the inflationary region for $\alpha \to \infty$.
\begin{table}\centering
   \begin{tabular}{|p{0.7cm}||p{1.7cm}|p{0.8cm}|p{1.7cm}|p{1.5cm}|}
 \hline
 $\alpha$ & $M_{min}[GeV]$ & $\phi^{min}_F$ & $T^{max}_{reh}[GeV]$& $T^*_{reh}[GeV]$\\ 
 \hline
  $ 10^8$    & $5.5\cdot 10^4$ & 4.01 & $1.2\cdot 10^6$&  $2.9\cdot 10^7$ \\
  $10^{12}$ & $1.7\cdot 10^5$ & 4.01 &  $1.5\cdot 10^2$&  $1.1\cdot 10^5$    \\
  $10^{16}$ & $5.5\cdot 10^5$ & 4.01 &  0.015          &  305      \\ 
  
 \hline
\end{tabular} 
\caption{Results for reheating for the $e$PV potential in eq.~\eqref{eq:peebles:exp} with $k=q=4$. The estimated maximum reheating temperature for the model $T^{max}_{reh}$ is below the lower bound $T^{*}_{reh}$ for every choice of $\alpha$, this implies that in general the model cannot account for the BBN constraints on overproduction of GWs (see appendix \ref{sec:appendix}). However compatibility with observations are restored if one assumes production of heavy particles in the early universe.}
\label{tab:peebles_k8_exp}
\end{table}
After inflation is over, kination and reheating follow (see sections \ref{sec:kination} - \ref{sec:reheating}). The reheating temperature must satisfy the bound $T_{reh}>T_{BBN}\sim 10^{-2}$ GeV. After reheating takes place, radiation eventually starts dominating. During the expansion, the scalar field $\phi_F$ freezes at the value given by eq.~\eqref{eq:frozen_scalar}. 

In order to have a working quintessential inflation mechanism, we need to satisfy two more requirements. The first one is that the value of the scalar field potential energy-density at freezing must match the value of the observed energy density of dark energy today (coincidence requirement). By using eqs.~\eqref{eq:tail_log} and $\eqref{eq:peebles:exp}$, we find:
\be\label{eq:coincidence_log}
U(\phi_F) \approx \frac{M^k}{4\alpha (\phi_F^k+\mu^k)} = \rho_0 \sim 7\cdot 10^{-121}.
\ee

Second, we need to check that the barotropic parameter at the present day $w_0$ for the quintessential tail is within the Planck observational bounds \cite{Planck:cosmo}. 
In order for this to happen we need to make sure that the scalar field unfreezes only at the time of matter-dark energy equality. In other words, the field stays frozen until recent times and only unfreezes when it becomes dominating. The field  then thaws and starts slow-rolling down its potential. This happens provided that the frozen scalar field does not hit the tracker before it starts dominating (see section \ref{sec:quintessence}). 
The equation for the barotropic parameter in $w_\phi$ is given by eq.~\eqref{eq:barotropic_phi}.
Once again the field will not substantially evolve till the present day, with $\phi_0\simeq \phi_F$ and $w_\phi \approx -1$.

We show in Table \ref{tab:peebles_k8_exp} the results for the case $k=q=4$. We only report the values of $\alpha$ for which we obtain viable CMB observables. The value $M_\text{min}$ is the minimum value of the mass scale for the potential in eq.~\eqref{eq:peebles:exp} necessary to avoid the tracker solution and let the field unfreeze only at matter-dark energy equality. The corresponding $\phi_F^\text{min}$ is obtained by imposing the solution of the coincidence problem in eq.~\eqref{eq:coincidence_log}. The reheating temperature is computed by means of eq.~\eqref{eq:T_reheating} by setting $M=M_\text{min}$ and $\phi=\phi_\text{min}$. Finally, we need to check if the model can satisfy the constraints on $T_{reh}$ coming from GW overproduction. During kination, the $w = 1$ stiff period induces a spike in the density of the GWs, potentially spoiling BBN. This can be avoided if $T_{reh}$ is large enough. The details on the calculation can be found in appendix \ref{sec:appendix}, while we show in Table \ref{tab:peebles_k8_exp} an estimate of the lower bound $T^*_{reh}$ (depending on the value of $\alpha$) that avoids GW overproduction.

Unfortunately, we always have $T_{reh}^{max}<T^*_{reh}$ hence the model is not in general viable. As before, compatibility with observation can be restored if we assume the gravitational production of very massive particles after the inflationary era \cite{saviour} which naively relaxes the reheating temperature bound to the one showed in eq. \eqref{eq:saviour}, making the results computed in Table \ref{tab:peebles_k8_exp} viable again.
\subsection{A note on the exponential tail}
Given eq.~\eqref{eq:tail_log}, one can try to reproduce an exponential tail by choosing a Jordan frame potential in the form of a double exponential:
\be
V(\phi) = 1- e^{\sigma e^{\lambda\phi}}.
\ee
The Einstein frame tail would then behave as:
\be
U(\phi)\approx \frac{e^{-\lambda\phi}}{4\alpha\lambda}.
\ee
However, this setup cannot be used to generate a quintessential tail. 

Consider the slow-roll parameter $\epsilon (\phi)= \frac{U'(\phi)^2}{2U(\phi)^2}$ with $U(\phi)$ given by eq.~\eqref{eq:einstein_potential_log}. It can be straightforwardly computed that this function is a monotonically increasing function with the following limits:
\begin{align}
  \epsilon(\phi) &\sim 0, & \text{for}&\hspace{3pt} \phi\to -\infty \, ,\\
 \epsilon(\phi) &\sim \frac{\lambda^2}{2}, & \text{for}&\hspace{3pt} \phi\to \infty \, ,
\end{align}
which implies that to end slow-roll during inflation and have a graceful exit we need $\lambda>\sqrt{2}$. 
Once again the same parameter determines the behavior at inflation and in the tail, this implies that the two scales cannot be decoupled.
Since this setup is equivalent to the one presented in \ref{sec:exponential}
we conclude that it is not viable.

\section{Conclusions}\label{sec:conclusions}

We have studied modeling  quintessential inflation in the context of $F(R,X)$ Palatini gravity. In particular, we considered a potential in the form of the generalized Peebles-Vilenkin (PV) potential for a quadratic $F(R,X)$. We proved that the model generates viable quintessential inflation in the case $k=q=4$ with a mass scale $M\sim 10^3-10^5$ GeV, providing the right inflationary observables, a solution for the coincidence problem and a prediction for the barotropic parameter $w_{\phi} \approx -1$. The model predicts in general a value of $T_{reh} < 10^5\,$GeV, which contradicts the lower bound on $T_{reh}$ necessary to avoid overproduction of GWs during kination. However, compatibility with observations can be restored if we assume production of heavy particles $\sim 10^{-6} m_P$, which later-on decay in the SM sector. This relaxes the bound to $1\,\text{MeV}\leq T_{reh}\leq 5\cdot 10^7\,$GeV, which is compatible with our predictions.

We also considered an example for a model $F(R,X)_{>2}$, in the form $F(R_X) = R_X + \alpha R_X^2 \ln(\alpha R_X)$ with a Jordan frame potential given by an exponential version of the PV potential, characterized by two mass scales $\mu, M$.
The model predicts viable quintessential inflation for $k=q=4$ with a mass scale of order $M \sim 10^5$ GeV which solves the coincidence problem and $1\ \text{MeV}<\mu< 10^3$ GeV fixed by the amplitude of scalar perturbations $A_s$. 
As in the previous case the model predicts in general a value of $T_{reh} < 10^5$ GeV which cannot be accepted if one considers the lower bound on $T_{reh}$ necessary to avoid overproduction of GWs during kination. The solution is again to assume production of heavy particles in the early Universe.

We also showed that a quintessential tail given by a simple exponential, although generating very good inflationary results, does not provide a solution for the coincidence problem (in both the $F(R,X)$ models). 

All in all, this study demonstrated that $F(R,X)$ Palatini gravity is a promising framework for constructing viable quintessential inflation models, although it can be challenging to address successfuly all the relevant constraints and, in particular, the overproduction of primordial gravitational waves during kination.

\section*{Acknowledgments}
KD is supported (in part) by the STFC consolidated Grant:
ST/X000621/1. 
This work was supported by the Estonian Research Council grants PRG1055,  RVTT3, RVTT7 and the CoE program TK202 ``Foundations of the Universe'’.  This article is based upon work from COST Actions COSMIC WISPers CA21106 and
CosmoVerse CA21136, supported by COST (European Cooperation in Science and Technology).

\appendix

\section{Constraints from the overproduction of GWs} \label{sec:appendix}

In the following we briefly compute an estimate on the reheating temperature necessary to avoid overproduction of GWs during kination (see for example \cite{Chen:2025awt} for the details).

In order to respect the bounds from BBN we need to constraint the intensity:
\be\label{eq:GW_integral}
\Omega_{peak}h^2 = \int^{\nu_{end}}_{\nu_{BBN}}\frac{\Omega_{GW}h^2}{\nu}  {d\nu}\leq \frac{7}{8}\qty(\frac{4}{11})^{4/3}\Omega h^2 \Delta{N_{\rm eff}}\,,
\ee
where $\Omega_{GW}$ is the spectrum of the gravitational waves, $\Omega_r h^2 \sim 2.47\cdot 10^{-5}$ is the relic density of radiation, $\nu_{BBN}\sim 10^{-11}$Hz and $\Delta N_{\rm eff} \sim 0.17$ the extra relativistic degrees of freedom during BBN given by the current Planck bound \cite{Planck:cosmo}. 

The above can be related to the reheating temperature as follows. Consider the frequencies $\nu_{end}$, $\nu_{reh}$ corresponding to the GW modes that reenter the horizon respectively at the end of inflation and at reheating. We have:
\bea
\nu_{end} &=& \frac{H_{end}}{2\pi}\frac{a_{end}}{a_0} \,\label{eq:nu_end} , \\
\nu_{reh}&=&\nu_{end}\qty(\frac{H_{end}}{H_{reh}})^{2/3} \,\label{eq:nu_reh} ,
\eea
with 
\be\label{eq:H_reh}
H_{reh} = \pi\sqrt{\frac{g^*_{reh}}{90}}T^2_{reh}.
\ee
Finally, between the end of the inflation and reheating we have that
\be\label{eq:Omega_P}
\Omega_{peak}=\Omega^{rd}_{GW}\frac{\nu_{end}}{\nu_{reh}} \, ,
\ee
where $\Omega^{rd}_{GW}\propto H_{end}^2$, is the GW density parameter of the modes that reenter the horizon during radiation domination.
By comparing \eqref{eq:GW_integral} with \eqref{eq:Omega_P} and using \eqref{eq:nu_end}-\eqref{eq:H_reh} we get a lower bound on $T_{reh}$. In this way, we can obtain the constraint on the reheating temperature appearing in Tables~\ref{tab:peebles_k4} and \ref{tab:peebles_k8_exp}.

\section*{Data Availability Statement}
No Data is associated with the manuscript.

\bibliographystyle{spphys}
\bibliography{references}

\end{document}